\documentclass{article}
\usepackage{mathtools}
\usepackage{amssymb}
\usepackage{tikz}
\usepackage{siunitx}
\usepackage{biblatex}

\usetikzlibrary{quantikz2}
\usepackage{graphicx}
\usepackage{caption}
\usepackage{subcaption}

\title{Experimental Approaches to Distinguishing Quantum Collapse from Unitary Evolution: A Weak Measurement Perspective}
\author{Peter Renkel}
\date{\today}
\begin{document}
\maketitle
\begin{abstract}

This paper proposes an experiment designed to distinguish between competing interpretations of quantum mechanics: those that involve wave function collapse and those that assume purely unitary evolution. The experiment tests whether an observer can measure a system without collapsing its wave function.

To this end, we introduce the concept of an unconscious observer, defined by two criteria:
(1) It measures a quantum system and sets the state of another system based on the result.
(2) It allows an external experimentalist to infer the measurement outcome by examining the observer’s state.
The more distinguishable the observer’s resulting states, the more it resembles a conventional measurement apparatus. Using weak measurements, the experimentalist probes these states, thereby testing the second criterion. 

The interference patterns observed in this setup reveal whether collapse has occurred, allowing experimental discrimination between collapse-based and unitary interpretations.

\end{abstract}
\section{Introduction}

Quantum mechanics (QM) underpins our understanding of the microscopic world, yet its interpretation remains a topic of ongoing debate. Various interpretations \cite{Copenhagen, Bohm, MWI, Histories, QBism} offer different perspectives on how the mathematical formalism of QM relates to physical reality.

This proposal outlines an experiment designed to differentiate between two broad categories of interpretations: those involving wavefunction collapse, such as the Copenhagen Interpretation \cite{Copenhagen}, and those based solely on unitary evolution, such as the Many-Worlds Interpretation \cite{MWI}. Earlier efforts to address this issue, including Deutsch’s thought experiments \cite{Deutsch}, have relied on idealized setups that were impractical with existing technology. In contrast, we propose a feasible, implementable experiment.

Traditionally, observers in QM have been construed as conscious agents interacting with the environment. Although the notion of unconscious observers has been briefly explored (e.g., in Everett’s original PhD thesis \cite{Everett}), the emphasis on consciousness has historically complicated experimental designs aimed at distinguishing interpretations.

In this study, we narrow the concept of an observer. Here, an \textbf{unconscious observer} is defined as a system that interacts with another system, extracts information about its state, and sets the state of a third system based on that information. By removing the requirement of consciousness, we sidestep interpretational issues tied to sentient observers. 

However, this definition alone is insufficient. For example, a CNOT quantum gate \cite{CNOT} sets the target qubit based on the control qubit’s state, yet it does not qualify as an observer in our sense. To refine the definition, we introduce an additional criterion: an unconscious observer must allow an external experimentalist to infer the outcome by examining the observer’s state. The more distinguishable these internal states are, the more the system behaves like a traditional measurement device. While the CNOT gate satisfies the first condition, it fails the second—it does not retain a record of the measurement outcome in its own state.

 To summarize, an unconscious observer must:
\begin{enumerate}
    \item Interact with a system, gather information about its state, and set the state of another system accordingly;
    \item Enable an external experimentalist to infer the outcome by examining the observer's state.
\end{enumerate}

The experimental setup proceeds as follows. A quantum system $S$ is measured by a candidate unconscious observer. The relevant measurement operator has two eigenvalues (0 and 1), with corresponding eigenvectors $\ket{0}$ and $\ket{1}$. Following this interaction, the candidate sets a second system, $S'$, to match the state of $S$.

We first evaluate whether this device qualifies as an unconscious observer. This is done via weak measurements that assess how effectively it distinguishes between outcomes.

Once a valid unconscious observer is confirmed, we apply a unitary rotation to the combined $S$–$S'$ system and measure the probability of finding $S'$ in the state $\ket{0}$. Collapse-based interpretations, such as the Copenhagen interpretation~\cite{Copenhagen}, predict this probability to be 0.5. In contrast, unitary-only interpretations predict a deviation from 0.5.

We then determine whether the observed probability is sufficient to statistically reject collapse-based interpretations. If the unconscious observer distinguishes outcomes very effectively, the interference effects are suppressed, and the deviation from 0.5 may be small, requiring high experimental sensitivity to detect it.

It is crucial to emphasize that the observer’s measurement and the subsequent unitary evolution are entirely automated—there is no conscious intervention. The external experimentalist only participates in the final verification step, where the outcome is assessed and potential interference is observed.

\section{Related Work and Existing Approaches}
The debate surrounding the interpretation of quantum mechanics, especially concerning the nature of measurement and wave function collapse, has been a central topic in physics for decades. Early theoretical analyses, such as Deutsch's \cite{Deutsch} influential thought experiments and Wigner's \cite{Wigner} seminal "Wigner's friend" scenario, provided crucial conceptual frameworks for exploring these distinctions. However, these foundational ideas often relied on highly idealized conditions or involved the ambiguous role of a "conscious observer," making experimental realization quite challenging.

More recently, experimental physics has pushed these boundaries, leading to concrete tests inspired by these thought experiments. Notably, experiments based on extended Wigner's friend setups \cite{Proietti, Bong} have begun to probe the very consistency of objective facts in quantum mechanics, often by demonstrating violations of certain Bell-like inequalities. While highly impactful, these experiments primarily focus on whether quantum mechanics allows for a shared, objective reality among multiple observers.

Our proposed experiment offers a distinct yet complementary approach. Instead of challenging the notion of objective outcomes themselves, we operate within a framework where objective measurement outcomes are assumed. Our goal is to directly test the mechanism by which these outcomes arise: specifically, to experimentally distinguish between interpretations that incorporate an explicit wave function collapse (like the Copenhagen interpretation \cite{Copenhagen} and various \textbf{objective collapse models} \cite{Ghirardi}, \cite{Pearle}, \cite{Penrose} that propose a fundamental physical process leading to collapse) and those based solely on purely unitary evolution (such as the Many-Worlds Interpretation \cite{MWI}, where observed classicality emerges from quantum entanglement and decoherence without an explicit collapse postulate).

A key implication of our approach is that \textbf{observing persistent interference, where collapse-based theories predict its absence, would constitute strong evidence against objective collapse of the wave function at the scale of our experimental apparatus.} This allows for a direct evaluation of just how "well-defined" the measurement performed by an "\textbf{unconscious observer}" needs to be for a full wave function collapse to occur. By probing the residual coherence of the system after its interaction with this unconscious observer, we can reveal interference effects that would be suppressed or absent if a complete, objective collapse had already taken place.

\section{Conceptual basis for testing the collapse-based interpretation using an unconscious observer}
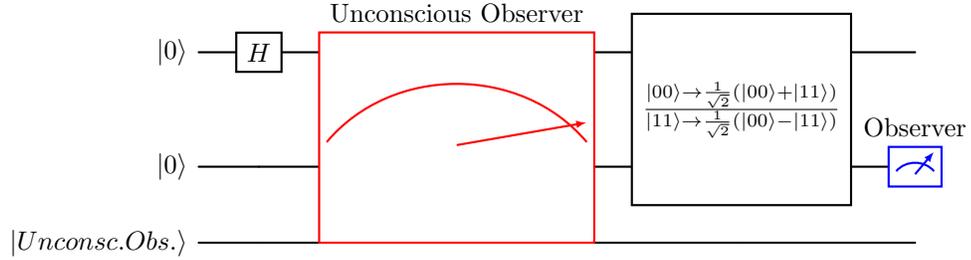
\begin{figure}
 \begin{quantikz}
\lstick{\ket{0}} &\gate{H} & \meter[4, style = {draw=red}]{\text{Unconscious Observer}}&\gate[2]
{
\frac{\ket{00} \to \frac{1}{\sqrt{2}}(\ket{00}+\ket{11})}
{\ket{11} \to \frac{1}{\sqrt{2}}(\ket{00}-\ket{11})}
}& \\
\lstick{\ket{0}} &&&&\meter[1, style = {draw=blue}]{\text{Observer}}\\
\lstick{\ket{Unconsc. Obs.}}&&&& \\
\end{quantikz}
\caption{Quantum circuit for the experiment aimed at distinguishing collapse-based from unitary-evolution-based interpretations. Central to this experiment is unconscious observer, which measures the state of the first qubit and prepares the second qubit accordingly.}
\label{fig:qcircuit}
\end{figure}

We commence our experimental procedure, as depicted in Fig.~\ref{fig:qcircuit}, by preparing two qubits in the state $\ket{00}$.  Subsequently, we apply a Hadamard gate \cite{Hadamard} to the first qubit, inducing a superposition of $\ket{0}$ and $\ket{1}$. Based on the measurement outcome, the unconscious observer sets the configuration of the second qubit.  
Specifically, if the measurement yields a result of 0, the second qubit remains unchanged; however, if the outcome is 1, the second qubit is set to $\ket{1}$.  We emphasize that the measurement apparatus operates autonomously, devoid of consciousness, yet akin to a conscious entity, it utilizes the measurement result to execute an action - the setting of the second qubit. Thus, our initial criterion, \textbf{the observer interacts with the system, gathers information about its state, and sets
another system’s state based on this information,} is met. Section \ref{sec:verification} addresses the second criterion.

We then apply a rotation transformation to the two qubits as follows:
\begin{equation} \label{rotation}
\begin{split}
\ket{00} &\to \frac{1}{\sqrt{2}}(\ket{00}+\ket{11}) \\
\ket{11} &\to \frac{1}{\sqrt{2}}(\ket{00}-\ket{11}) \\
\ket{01} &\to \ket{01} \\
\ket{10} &\to \ket{10} \\
\end{split}
\end{equation}
and measure the value of the second qubit.

\section{Collapse-based interpretations}
We will use the Copenhagen interpretation as a representative example of collapse-based interpretations, as the results for most other collapse-based interpretations are expected to be identical. According to this interpretation, the act of measurement by the unconscious observer collapses the state of the first qubit to either $\ket{0}$ or $\ket{1}$ with equal probability $\frac{1}{2}$. In the former scenario, the system's state will be $\ket{00}$ after setting the second qubit to $\ket{0}$. In accordance with Equation \ref{rotation}, following the rotation, it evolves into $\frac{1}{\sqrt{2}}(\ket{00} + \ket{11})$. Conversely, in the latter case, the system's state will be $|11\rangle$, and after the rotation, it transforms into $\frac{1}{\sqrt{2}}(|00\rangle - |11\rangle)$.

Using Eq.\ref{rotation} and defining the projection operator on the $\ket{0}$ state of the second qubit as:
\begin{equation}
    \hat{P}_2^0 = \ket{00}\bra{00} + \ket{10}\bra{10},
\end{equation}
the probability of observing the second qubit in the state $\ket{0}$ post-rotation is given by:

\begin{equation} \label{Copenhagen}
\begin{split}
p_2(0)_{\text{rot}}^{Copenhagen} = 
p_1(0) \times \left| \frac{1}{\sqrt{2}} \hat{P}_2^0 ( \ket{00} + \ket{11})  \right|^2 + \\ p_1(1) \times \left| \frac{1}{\sqrt{2}}\hat{P}_2^0 ( \ket{00} - \ket{11}) \right|^2 = \\
\frac{1}{2} p_1(0)  + \frac{1}{2} p_1(1),
\end{split}
\end{equation}

where $p_1(0)$ and $p_1(1)$ represent the probabilities of finding the first qubit in states $\ket{0}$ and $\ket{1}$, respectively. Since $p_1(0) = p_1(1) = \frac{1}{2}$,
\begin{equation}\label{res_Copenhagen}
p_2(0)_{\text{rot}}^{Copenhagen} = \frac{1}{2}
\end{equation}

\section{Unitary evolution interpretations}

Within unitary evaluation conceptualization, the unconscious observer ceases to be a mere observer but instead becomes an active participant in the process.

Let's denote the initial state of the unconscious observer as $\ket{obs_0}$. The state of the entire system before the measurement by the unconscious observer is performed is given by:
\begin{equation}
\ket{\psi} = \frac{1}{\sqrt{2}}(\ket{00} + \ket{11})\ket{obs}
\end{equation}
and following the measurement, it evolves to:
\begin{equation}
\ket{\psi'} = \frac{1}{\sqrt{2}}(\ket{00}\ket{obs'_{00}} + \ket{11}\ket{obs'_{11}})
\end{equation}
where $\ket{obs'_{00}}$ and $\ket{obs'_{11}}$ represent the states of the unconscious observer post-measurement. The unconscious observer measures the first qubit, thereby multiplying its eigenstates $\ket{0}$ and $\ket{1}$ by a constant. This constant is a pure phase since the unconscious observer's measurement can be represented as a unitary operator acting on the pre-measurement state. However, the observer alters its own state, and we incorporate this phase into it. 

After the rotation and utilizing Eq. \ref{rotation}, the state $\ket{\psi'}$ transforms as follows:
\begin{equation}\label{wf}
\begin{split}
\ket{\psi'} \rightarrow \ket{\psi''} = \frac{1}{\sqrt{2}}
(\frac{1}{\sqrt{2}}(\ket{00}+\ket{11})\ket{obs'_{00}} + \frac{1}{\sqrt{2}}(\ket{00}-\ket{11})\ket{obs'_{11}}) = \\
\frac{1}{2} (\ket{00} (\ket{obs'_{00}} +\ket{obs'_{11}}) + \ket{11} (\ket{obs'_{00}} -\ket{obs'_{11}}))
\end{split}
\end{equation}

After the final measurement, rendering the second qubit in state $\ket{0}$, $\ket{\psi''}$ transforms into
\begin{equation}
\begin{split}
\ket{\psi''} \rightarrow \ket{\psi'''} = \hat{P}_2^0\ket{\psi''} = \\(\ket{00}\bra{00} + \ket{10}\bra{10})\psi''=\\
\frac{1}{2} (\ket{obs'_{00}} +\ket{obs'_{11}}).
\end{split}
\end{equation}
The probability to find the second qubit in the state $\ket{0}$ after the measurement is:
\begin{equation}  \label{MWI}
p_2(0)_{rot}^{U} = \braket{\psi'''}{\psi'''} = \frac{1}{4} (2 +\braket{obs'_0}{obs'_1} + \braket{obs'_1} {obs'_0}) = \\
\frac{1}{2} (1 + Re(\braket{obs'_0}{obs'_1}))
\end{equation}

This result differs from Eq. \ref{res_Copenhagen} by an additional term
\begin{equation} \label{scalarproduct}
s = \frac{1}{2}Re(\braket{obs'_0}{obs'_1})
\end{equation}
Therefore, provided that \textbf{the scalar product of the observer's states corresponding to two measurement outcomes can be assessed} and \textbf {the scalar product of the observer's states corresponding to two measurement outcomes does not change significantly from measurement to measurement}, we will be able to distinguish the Unitarity based from the Copenhagen interpretation by repeating the experiment many times.

An experimentalist performing $N$ experiments should expect to find the second qubit in state $\ket{0}$ approximately $\frac{N}{2}$ times under the Copenhagen interpretation and $\frac{N}{2} + sN$ times under the Unitary based interpretations.

To reject the null hypothesis (Copenhagen), we need to establish a certain $p$-value. For instance, with $N$ measurements and a $p$-value of 95\%, we will need $sN$ to be twice as large as the standard deviation $\sigma=\frac{1}{2}\sqrt{N_{p=0.05}}$.

Hence,
\begin{equation} \label{Ns}
sN_{p=0.05} = \sqrt{N_{p=0.05}}
\end{equation}
and
\begin{equation} \label{Ntrials}
N_{p=0.05} = \frac{1}{s^2}.
\end{equation}
For $s<<1$, many measurements will be necessary to achieve our goal.

\section{Additional challenges to the experiment.}
Experimentalists must confirm that the measurement was conducted - the second criterion for the unconscious observer. 

Crucially, this verification should occur before measuring the second qubit, as the act of verification could alter the observer's state. It should also be a weak measurement to preserve the interference pattern. See Section \ref{sec:example} for a concrete example of performing such a verification.

A fundamental question arises regarding when a quantum measurement truly occurs. One could argue that an unconscious observer, as defined above, doesn't perform a complete measurement. Instead, the measurement is only finalized when a human observer interacts with the quantum observer's output, confirming whether the measurement took place. This viewpoint challenges the objectivity of measurement in quantum mechanics, suggesting that no measurement is fully realized until it is perceived by a conscious observer—raising questions about the existence of an external reality independent of observation.

Interactions with the environment—including photon emissions from the ions, whether spontaneous or stimulated—further complicate the situation by inducing decoherence in the system under observation. To maintain coherence, both the preparation and measurement of the second qubit must be completed before the environment fully decoheres the qubits and the measuring apparatus (see Appendix A for details).

Finally, no quantum gates and no unconscious observer are perfect, and errors may occasionally occur. This topic is discussed further in Appendix B.

\section{A concrete experimental realization} \label{sec:example}

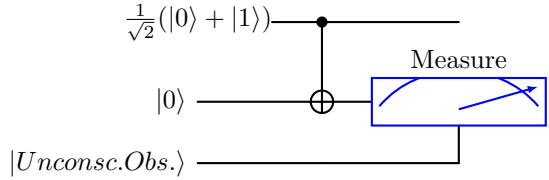
\begin{figure}
 \begin{quantikz}
\frac{1}{\sqrt{2}}(\ket{0} + \ket{1}) & \ctrl{1}&\\
\lstick{\ket{0}}&\targ{x}&\meter[1, style = {draw=blue}]{\text{Measure}}\ctrl{1}\\
\lstick{\ket{Unconsc. Obs.}}&& \\
\end{quantikz}
\caption{An example of an unconscious observer (highlighted in red in Fig.~\ref{fig:qcircuit}), which measures the first qubit and sets the second qubit accordingly. Its initial state is denoted as $\ket{\text{Unconsc.~Obs.}}$.}
\label{fig:obs_example}
\end{figure}

Figure \ref{fig:obs_example} illustrates how a quantum unconscious observer, satisfying the aforementioned requirements and indicated in red in Figure \ref{fig:qcircuit}, might be represented.

The three lines to the left correspond to the three inputs to the Unconscious Observer gate in Figure \ref{fig:qcircuit}. We apply a CNOT gate, bringing the system to the state: $\frac{1}{\sqrt{2}}(\ket{00} + \ket{11})$, then the unconscious observer measures the first qubit.

\begin{figure}[h]
\includegraphics[scale=0.5]{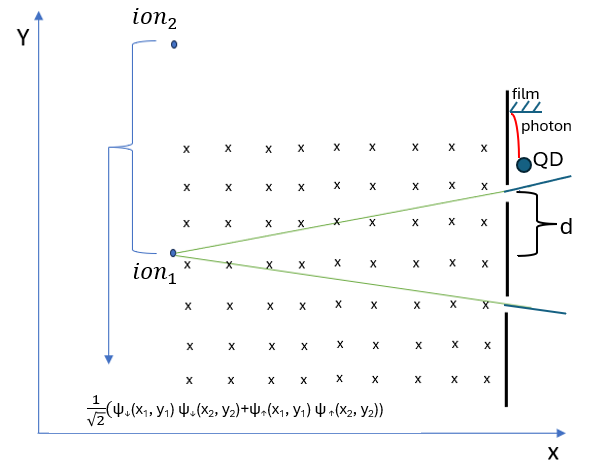}
\caption{A quantum dot (QD) acts as an unconscious observer, determining the spin of an entangled ion. The paths of the first ion are shown in green, with blue segments indicating measurement regions. A photon in the top right of the figure is used to read out the measurement result performed by the QD on the ion (one of the photon's possible paths is also shown).}
\label{fig:unconsc_obs}
\end{figure}

Figure \ref{fig:unconsc_obs} depicts a specific implementation of this concept. 

We begin by trapping and entangling two ions \cite{tions} (e.g. $Yb^+$, known for their long coherence times \cite{Nop2021YbIonTraps}). The two qubits in the above discussion are represented by the two entangled ions. 

The first ion is then directed through an analog of a double-slit experiment comprising two transport channels \cite{tchannels} with a magnetic field, $\mathbf{H}$, applied. A Quantum Dot (QD) is positioned near the end of the upper channel to serve as a measurement device.

We will work in the coordinate representation for the remainder of this section. The computational basis state \( \ket{0} \) corresponds to the spatial wavefunction \( \psi_{\downarrow}(x, y) \), where \( \downarrow \) denotes (nuclear) spin-down, and \( \ket{1} \) corresponds to \( \psi_{\uparrow}(x, y) \), where \( \uparrow \) denotes spin-up.

For practical purposes, the ions can be treated as distinguishable. This can be achieved either by preparing them in different internal states or by using two different isotopes of the same trapped ion species.

The wavefunction of the two-ion system after the application of a CNOT gate is given by:

\begin{equation}
\begin{split}
\ket{\psi} = \frac{1}{\sqrt{2}} \big( & \ket{\psi_{\downarrow}(x_1, y_1)} \ket{\psi_{\downarrow}(x_2, y_2)} \\
+ & \ket{\psi_{\uparrow}(x_1, y_1)} \ket{\psi_{\uparrow}(x_2, y_2)} \big), \label{initial_wf}
\end{split}
\end{equation}

where indices \( 1 \) and \( 2 \) denote the coordinates of the first and second ion, respectively.

The magnetic field directs the first ion along two paths simultaneously: the upper path for the $\uparrow$ component and the lower path for the $\downarrow$ component. We then need to transport the ions back for quantum computations preserving coherence. Such transportation might present significant practical challenges due to ion loss, control precision, and decoherence, but should be possible.

The QD near the upper path effectively measures the ion's spin (the extent to which this QD qualifies as an observer is described in Section~ \ref{sec:verification}). Due to the interaction with the QD, the ion 'chooses' a path. If it passes near the QD,  its spin is $\uparrow$, otherwise it is $\downarrow$. The entangled ion's spin is immediately set to the same value. See also \cite{Heiblum} and \cite{DoubleSlit1} for a similar experiment with electrons.

\subsection{Collapse-based interpretations}
The wave function of the pre-measurement two-ion system is given by:
\begin{equation} \label{QDsystem_copenhagen}
\begin{split}
\psi^{ions}_{pre-meas}(x_1, y_1, x_2, y_2) =  \\
\frac{1}{\sqrt{2}}(\psi_{\downarrow}(x_1)\delta(y_1+d)\psi_{\downarrow}(x_2, y_2) + \\
\psi_{\uparrow}(x_1)\delta(y_1-d)\psi_{\uparrow}(x_2, y_2)),
\end{split}
\end{equation}
 where $d$ is the distance from the center of the double-slit plate to each slit. The delta functions in the expression above ensure that, after passing through the slits, the first ion cannot be found anywhere other than directly in front of one of the slits.

The QD measures the $y$-component of the wave function of the first ion. This measurement collapses the wave function Eq. \ref{QDsystem_copenhagen} such that only one term survives: either corresponding to the ion passing through the upper slit and QD, or to the ion passing through the lower slit.

The wave function of the system of the two ions then becomes either
\newline
$\psi_{\downarrow}(x_1)\delta(y_1+d)\psi_{\downarrow}(x_2, y_2))$ or $
\psi_{\uparrow}(x_1)\delta(y_1-d)\psi_{\uparrow}(x_2, y_2)$. 

\subsection{Unitary evolution interpretations}
A quantum dot can be approximated as a quantum well. Let's expand the wave function of the first ion over the eigenfunctions of this potential well.
The wave function of the system of the two ions before the measurement then becomes:

\begin{equation} \label{QDsystem_wf}
\begin{split}
\psi^{ions}_{pre-meas}(x_1, y_1, x_2, y_2) =  \\
\frac{1}{\sqrt{2}}(\psi_{\downarrow}(x_1)\delta(y_1+d)\psi_{\downarrow}(x_2, y_2) + \sum_i \\
C^i \psi^i(x_1)\delta(y_1-d)\psi_{\uparrow}(x_2, y_2))),
\end{split}
\end{equation}
where $\psi^i(x_1)$ are the stationary eigenfunctions of the well and $C^i$ are the coefficients of expansion of $\psi_{\uparrow}(x_1)$ around these eigenfunctions. $i$ is continuous if ion's energy is above the maximum energy for the potential well, so the sum above can be replaced with an integral for a concrete potential. For now, we keep it the general form.

The wave function of the ions+QD system before the measurement is:
\begin{equation} \label{QDsystem_wf}
\begin{split}
\psi^{ions+QD}_{pre-meas}(x_1, y_1, x_2, y_2) = \psi^{ions}_{pre-meas}(x_1, y_1, x_2, y_2)\psi^{QD}(\Omega),
\end{split}
\end{equation}
where $\psi^{QD}$ is a wave function of the QD, and $\Omega$ denotes its all internal degrees of freedom.

Let's assume that the QD acts only on the coordinates of the first ion, and consider the QD-spin interaction as negligibly small. The QD measures the $y$ coordinate of the ion and leaves the delta function $\delta(y-y_0)$ intact.

The action of the unitary operator on the QD-ion system, when the first ion’s wave function is \( \psi^i(x) \) and the QD’s wave function is \( \psi^{QD} \), is given by the following equation:

\begin{equation}
\widehat{U} \delta(y - y_0) \psi^i(x) \psi^{QD} = \delta(y - y_0) \psi^i(x) \psi'^{QD, i},
\end{equation}
where \( \psi'^{QD, i}(\Omega) \) represents the wave function of the quantum dot after its interaction with the ion, with the ion's state equals to the \( i^{\text{th}} \) eigenfunction of the quantum well.

The wave function of the system of the two ions and the QD after the interaction with the QD becomes:

\begin{equation} \label{qdwf}
\begin{split}
\psi'^{ions+QD}_{post-meas}(x_1, y_1, x_2, y_2) = \\
\\ \frac{1}{\sqrt{2}}(\psi_{\downarrow}(x_1)\delta(y_1+d)\psi_{\downarrow}(x_2, y_2)\psi^{QD}(\Omega) + \\
\sum_i C^i \psi^i(x_1)\psi'^{QD, i}(\Omega)\delta(y_1-d) \psi_{\uparrow}(x_2, y_2))).
\end{split}
\end{equation}

Multiplying by $\psi^{*QD}$ and integrating over the internal degrees of freedom of the QD, we get:

\begin{equation}
\begin{split}
\int d\Omega \psi^{*QD}(\Omega)\psi'^{ions+QD}_{post-meas}(x_1, y_1, x_2, y_2) = \\
\frac{1}{\sqrt{2}}(\psi_{\downarrow}(x_1)\delta(y_1+d)\psi_{\downarrow}(x_2, y_2) +\\
\sum_i C^i\psi^i(x_1)
\int d\Omega\psi^{*QD}(\Omega) \psi'^{QD, i}(\Omega)\delta(y_1-d) \psi_{\uparrow}(x_2, y_2))),
\end{split}
\end{equation}
where we used the fact that $\int \psi^{*QD}(\Omega) \psi^{QD}(\Omega) d\Omega = 1$.

Denoting

\begin{equation}\label{scalar_products}
D^i = \int d\Omega \psi^{*QD}(\Omega) \psi'^{QD, i}(\Omega),
\end{equation}

we get:

\begin{equation}\label{el_after_interaction}
\begin{split}
\int d\Omega \psi^{*QD}\psi'^{ions+QD}_{post-meas}(x_1, y_1, x_2, y_2) = \\
\frac{1}{\sqrt{2}}(\psi_{\downarrow}(x_1)\delta(y_1+d)\psi_{\downarrow}(x_2, y_2) + \\
\sum_i D^i C^i\psi^i(x_1)
\delta(y_1-d) \psi_{\uparrow}(x_2, y_2))).
\end{split}
\end{equation} 
Here, the coefficients 
$D_i$
  depend on the quantum dot's interaction with the first ion. Notice, that we now eliminated the QD wave function entirely in the right handside of the above equation.

The presence of many coefficients \( C_i \) and \( D_i \) complicates both theoretical calculations and experimental setups. To simplify this, we propose preparing the spin-up state \( \psi_{\uparrow} \) as a narrow Gaussian centered around a selected eigenstate \( \psi^{i_0} \). We approximate the sum over \( i \) as a delta symbol(or delta function for the continuous spectrum), and therefore \( D^i \approx D \delta_{i, i_0} \) and \( \psi^{i_0}(x_1) \approx \psi_{\uparrow}(x_1) \).

Strictly speaking, any eigenfunction of the potential well is stationary and can't be used to describe a moving particle. Therefore, the impact of this assumption should be further evaluated for specific quantum dots.

The Eq. \ref{el_after_interaction} then becomes:
\begin{equation}\label{ion_after_interaction}
\begin{split}
\int d\Omega \psi^{*QD}\psi'(x_1, y_1, x_2, y_2)^{ions+QD}_{post-meas} = \\
\frac{1}{\sqrt{2}}(\psi_{\downarrow}(x_1)\delta(y_1+d)\psi_{\downarrow}(x_2, y_2) +\\
D \psi_{\uparrow}(x_1)
\delta(y_1-d) \psi_{\uparrow}(x_2, y_2)).
\end{split}
\end{equation} 

We propose waiting after each measurement and subsequent rotation, for the quantum dot (QD) to decohere, allowing its state to evolve into a stable pointer state of the environment. If the QD does not change its observable properties (e.g., the photons it scatters), its state should remain approximately the same as before the measurement. In other words, we assume the following process:
\begin{equation} \label{stable_scalarproduct}
\psi^{QD}(\Omega)\xrightarrow{measurement} \psi'^{QD}(\Omega)\xrightarrow{decoherence} \psi^{QD}(\Omega).
\end{equation}

Under this assumption, the coefficient $D$ remains relatively constant because $\psi^{QD}(\Omega)$ is always the same before each measurement. We want to stress that, although this assumption seems reasonable, it should be validated during the experiment.

Experiments, measuring $D$ for electrons, were performed in the past (see \cite{Heiblum}). It was shown that both its absolute value and phase were stable and didn't change from an experiment to an experiment. Analogous experiments for ions interacting with QD were not performed as far as we know.

By performing calculations similar to those used to derive Eq.\ref{MWI}, we obtain:

\begin{equation} \label{pD}
p_2(\downarrow) = \frac{1}{2}(1 + \text{Re}(D)).
\end{equation}

The probability differs from $0.5$ by a term:
\begin{equation} \label{interference}
    \frac{1}{2} Re(D)
\end{equation}
This term is responsible for the interference.
We need to perform $N > \frac{4}{Re(D)^2}$ (see Eq. \ref{Ntrials}) to distinguish between the interpretations with 95\% confidence.

\subsection{Verifying unconscious observer performed a measurement}

\label{sec:verification}

After interacting with the first ion, the quantum dot (QD) sets the second ion into the same spin state, thereby fulfilling the first requirement for an unconscious observer: \textit{it interacts with the system, gathers information about its state, and uses that information to influence the state of another system}.

To meet the second criterion—\textit{the observer must allow an external experimentalist to infer the measurement outcome by examining the observer’s state}—the experimentalist must probe the state of the QD after its interaction with the ion. This probing should occur immediately following the QD–ion interaction, before any additional operations (such as the ion rotation represented by the large black square in Fig.~\ref{fig:qcircuit}) are applied, as those could complicate the inference.

Crucially, this probing must be performed via a weak measurement, rather than a strong (projective) one, to preserve quantum coherence necessary for interference.

Therefore, to fulfill the second criterion for unconscious observation, the external experimentalist should perform a weak measurement of the QD's state immediately after interacting with the first ion.

We aim to assess the likelihood that our measurement apparatus successfully performed a measurement. To do this, we direct a single photon—generated, for instance, by an excited atom—toward the original position of the quantum dot (QD) and detect its location on the film (\ref{laser_a} and \ref{laser_b}). The state of the system, comprising the ions and QD after the photon hits the screen can be written as:

\begin{figure}
\centering
\subfloat[A directed single photon hits the film. QD is shifted.]{
\includegraphics[scale=0.45]{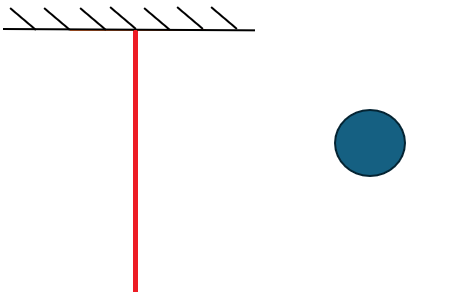}
\label{laser_a}}
\hspace{.125\textwidth}%
\subfloat[A QD scatters the photon, so the probability to find it further from the original point is higher]{
\includegraphics[scale=0.45]{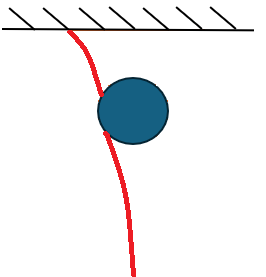}\label{laser_b}}
\caption{Assessing the state of the QD after its interaction with the first ion using photon scattering.}
\end{figure}
\begin{equation}
    \begin{split}
\ket{sys}^{\text{measured}} = F(r)\ket{QD''} \ket{\downarrow}_1\ket{\downarrow}_2 + F'(r)\ket{QD'''}\ket{\uparrow}_1\ket{\uparrow}_2 ,
\end{split}
\end{equation}
where $\ket{QD''}$ is 
 the state of the QD after scattering the photon, $\ket{QD'''}$ is the state of QD after interacting with both the photon and the ion, $\ket{\downarrow}_{1, 2}$ represent the states of the first and second ions with spin down, and $\ket{\uparrow}_{1, 2}$ represent the spin-up states.
The complex-valued coefficients $F(r)$ and $F'(r)$ can be numerically calculated from the Mie scattering theory for a concrete realization of the QD and the photon emitters \cite{Mie}),
\begin{equation}
    |F(r)|^2 + |F'(r)|^2 = 1,
\end{equation}
where the polar radius, $r$, is defined as the distance between the point of photon impact on the film and the projected center of the quantum dot before its interaction with the photon. This projection is obtained by mapping the geometric center of the quantum dot onto the plane. 

If  $|F'|\sim 1$ and $|F|\sim 0$, the measurement indicated that the ion passed close to the QD. 
If $|F'|\sim 0$ and $|F|\sim 1$, the ion chose another path. 
Finally, if neither of the above conditions holds, the QD (coupled with a photon and a film) couldn't identify the path ion took.  Let's introduce discriminability, the measurement strength:
\begin{equation}
\Lambda = \frac{1}{|F(r)F'(r)|} \label{classicity}
\end{equation}
The discriminability indicates how well our QD discriminates between the ion's paths. More concretely, it represents the inverse of the square root of the product of probabilities to find the QD in each of the two states: $\ket{QD''}$ or \ket{QD'''} if we follow our weak measurement with a strong measurement.

We can assume the QD a measuring device if 
\begin{equation}
\Lambda>\Lambda_0 = 4.6 \label{Lambda},
\end{equation} 
where 4.6 corresponds to the probabilities of 95\% and 5\%.

As we know, the fact of measurement, perturbs the system (see \cite{Busch}). Let's understand how our system is affected by the weak measurement.

After the ion rotation (Eq. \ref{rotation}), we get:
\begin{equation}
    \begin{split}
    \ket{sys}^{rotation} = 
    \frac{1}{\sqrt{2}}(\ket{\downarrow}_1\ket{\downarrow}_2(\ket{QD''}F(r) + \ket{QD'''}F'(r)) + \\
    \ket{\uparrow}_1\ket{\uparrow}_2(\ket{QD''}F(r) - \ket{QD'''}F'(r))
    \end{split}
\end{equation}

Hence, the probability of finding the second ion in the spin-down state is given by:
\begin{equation}
\begin{split}
\label{final_interference}
    p_2(\downarrow) = 
    \frac{1}{2}\left(1 +  \text{Re}\left[F^*(r)F'(r)D'\right]\right) = 
    \frac{1}{2}\left(1 +  \frac{\text{Re}(D')}{\Lambda} \cdot \frac{\cos(\phi_F + \phi_{D'})}{\cos(\phi_{D'})} \right),
\end{split}
\end{equation}
where
\begin{equation}
\label{scalar_products1}
D' = \braket{\text{QD}''}{\text{QD}'''} = \int d\Omega\, \psi^{*}_{\text{QD}''}(\Omega)\, \psi'^{\text{QD}''', i_0}(\Omega).
\end{equation}

This result modifies the interference term in Eq.~\ref{interference} by a factor of 
$$
\frac{\text{Re}(D')}{\text{Re}(D)} \cdot \frac{1}{\Lambda} \cdot \frac{\cos(\phi_F + \phi_{D'})}{\cos(\phi_{D'})}.
$$
Although this factor depends on the phases of $D'$ (denoted $\phi_{D'}$) and of $F(r)F'(r)$ (denoted $\phi_F$), as well as on the ratio $\text{Re}(D')/\text{Re}(D)$, we expect the dominant contribution to arise from the $1/\Lambda$ factor. By design, this factor is less than $1/4.6 \approx 0.2$, implying that the interference effect is reduced approximately by a factor of $\Lambda$.

This gives us a tool to test (reject or confirm with some p-value) the collapsed-based interpretations, if we collect statistics for all measurements, for which Eq. \ref{Lambda} holds, and still discover the interference. For example, for $Re(D)$ of 0.1, we will need approximately $2000$ measurements.

\section{Extending the experiment}

 Instead of entangling the first ion with the second one, we can \textbf{condition the state of the second ion based on the measurement outcome of the first ion}. In this scenario, the photon’s position gives insight into the path taken by the ion. If the first ion passes near the QD, it shifts, and the likelihood of detecting the photon closer to the center of the film increases.
 Conversely, if the first ion takes the lower path, the QD doesn't shift, and the likelihood of detecting the photon far to the center of the film increases.

We start with preparing the second ion with its spin down. We then flip its spin conditionally, depending on whether we detect the photon near the QD center. A photon presence within a circle where $|F'|/|F|=\sqrt{\frac{95\%}{5\%}} = 4.4$ or in a ring $|F|/|F'| = 4.4$ will serve as the conditional qubit for the CNOT gate.
We will ignore interactions when both $|F'|/|F| < 19$ and $|F|/|F'| < 19$. One way to do it is to automatically set the flag to 1 for interactions that satisfy our unconscious observer criteria in some register and collect statistics only for those. 

This method automatically prepares the state $\ket{\uparrow}\ket{\uparrow}$ when the quantum dot (QD) registers the first ion taking the upper path, and $\ket{\downarrow}\ket{\downarrow}$ for the lower path. The persistence of interference under these conditions would challenge collapse-based interpretations, specifically for observers with discriminability exceeding the threshold $\Lambda > \Lambda_0$.

Furthermore, using Eq. \ref{final_interference} and utilizing $F(r)$ coefficients, we can compute $D'$. This will enable us to test Eq. \ref{final_interference} for different scattering arrangements (e.g. varying the distance between the QD and the above-mentioned circle or varying the wavelength of incoming photon). If confirmed, this would decisively demonstrate that quantum evolution in measurement remains strictly unitary, thereby challenging interpretations that predict deviations due to collapse mechanisms.

\section{Summary}

We have demonstrated that an experiment capable of distinguishing between collapse-based and unitary interpretations of quantum mechanics is, in principle, feasible. Section~\ref{sec:example} outlines key theoretical assumptions for experimental validation, while Appendices A–C address practical considerations such as environmental interactions and their impact on the experiment's viability.

This experiment utilizes an unconscious observer, defined by the following criteria:

\begin{enumerate} \item It interacts with the system, gathers information about its state, and sets the state of another system based on this information. \item Its internal state must encode the measurement outcome in a manner accessible to an external experimentalist. \end{enumerate}

To successfully implement such an experiment, the following conditions must be satisfied:

\begin{itemize} \item \textbf{Minimized Environmental Interaction:} The observed system must remain largely isolated from its environment. This minimizes decoherence and preserves quantum coherence, ensuring that interference effects are not obscured (see Eq.~\ref{environment}).
However, to test whether collapse requires environmental interaction, the experiment may additionally be performed under conditions of weak environment coupling (see Appendix A).

\item \textbf{Measurable Interference:} The observer’s resulting states—corresponding to different measurement outcomes—should be nearly orthogonal. This small overlap enables the detection of quantum interference between the outcomes (see Eq.~\ref{final_interference}).

\item \textbf{Stable Observer States:} The inner product between observer states corresponding to different outcomes should remain approximately constant across repeated trials. This ensures the reliability and reproducibility of interference measurements (see Eq.~\ref{stable_scalarproduct}). \end{itemize}

We present a concrete realization of an unconscious observer in which all the above conditions are met, the second criterion is achieved by utilizing single photons and the first criterion - via entanglement between the measured system and the observer. We further propose an extension in which the observer sets the secondary system's state using classical algorithms based on the measurement result. 

\section*{Acknowledgement}

I am particularly grateful to Vadim Asnin for invaluable discussions, providing feedback on drafts, and proofreading. His contributions significantly improved the quality of this paper.

\printbibliography

\section*{Appendix A. Environment}

If the interaction with the environment were negligibly small, the density matrix representing the system after the rotation, in the unitary evolution picture, would be given by $\ket{\psi''}\bra{\psi''}$. Using Eq.~\ref{wf}, the density matrix $\rho$ in the computational basis becomes:
\begin{equation} \label{dmbefore}
\rho =
\begin{pmatrix}
\frac{1}{2} \left(1 + \text{Re}(\braket{obs'_0}{obs'_1})\right) & i\frac{1}{2} \text{Im}(\braket{obs'_0}{obs'_1})\\
-i\frac{1}{2} \text{Im}(\braket{obs'_0}{obs'_1}) & \frac{1}{2} \left(1 - \text{Re}(\braket{obs'_0}{obs'_1})\right)
\end{pmatrix},
\end{equation}
in the $\ket{00}$, $\ket{11}$ basis. The element $\rho_{00,00}$ defines the probability of finding the system in the state $\ket{00}$ (see Eq.~\ref{MWI}).

If the interaction with the environment is not negligibly small, the system will undergo decoherence. 
Following the assumptions in~\cite{PointerStates}, we postulate that the environment interacts with the system exclusively after the measurement.
After decoherence, the density matrix becomes diagonal in the pointer basis~\cite{Zurek}. Whether this pointer basis coincides with the computational basis $\{\ket{00}, \ket{01}, \ket{10}, \ket{11}\}$ depends on the specific form of the interaction between the environment and the ions.

To rigorously determine the measurement probabilities after decoherence and an additional rotation, one must solve the master equation for the specific system–environment interaction Hamiltonian. For illustrative purposes, we assume that the pointer basis coincides with the computational basis.

Under this assumption, the density matrix after decoherence takes the form:
\begin{equation}
\rho_{\text{decoherence}} =
\begin{pmatrix}
\frac{1}{2} & 0 \\
0 & \frac{1}{2}
\end{pmatrix}.
\end{equation}

As a result, the measurement probability becomes $\frac{1}{2}$, and it is no longer possible to distinguish between the two interpretations. The detailed time evolution of $\rho_{00}(t)$ depends on the nature of the environment and its coupling to the system.

 Experimentalists must ensure that the final measurement takes place before decoherence sets in and that the required number of experiments for significance level 0.05, $N_{p=0.05}$, is experimentally feasible. This requires isolating the system from environmental noise and ensuring that all rotations and measurements occur on timescales much shorter than the decoherence time. Additionally, to prevent spontaneous emission, they must avoid resonances with internal QD transitions. This ensures that the QD's recoil remains coherent and does not leak which-path information to the environment. 
 
 But to rule out the hypothesis that there is no collapse without interaction with the environment, a controlled interaction with the environment might be permitted. This interaction should be modelled with the master equation in the context of a concrete experiment to update predictions given by Eq. \ref{final_interference}.  The precise form of this interaction depends on the experimental implementation and is beyond the scope of the present study. For an evaluation of environment interactions in the double-slit framework, see \cite{WhichPath}.

Analogously to Eq.~\ref{Ntrials}, the number of measurements $N$ required to distinguish between interpretations, taking the environment into account, can be estimated as:
\begin{equation} \label{environment}
N_{p=0.05} = \frac{1}{\left(\rho_{00}(t_{\text{meas}}) - \frac{1}{2}\right)^2},
\end{equation}
where $t_{\text{meas}}$ denotes the time at which the final measurement of the second qubit occurs.

Furthermore, as a consistency check, experimentalists can perform several sets of $N$ measurements, each set corresponding to a different measurement time $t_{\text{meas}}$. By profiling the dependence of $N \rho_{00}(t_{\text{meas}})$ on $t_{\text{meas}}$, they can test the interpretation: if the unitary evolution holds, $N \rho_{00}(t_{\text{meas}})$ will start above or below $\frac{N}{2}$ and asymptotically approach $\frac{N}{2}$ with time. In contrast, under a collapse-based interpretation, $N \rho_{00}(t_{\text{meas}})$ will remain at $\frac{N}{2}$ at all times.

\section*{Appendix B. Measurement errors}

All errors can be broadly divided into two categories: those associated with quantum gates, and those arising from the unconscious observer.

Imperfect quantum gates—such as a non-ideal Hadamard gate—can introduce systematic errors into the measurement outcomes. As a result, the probability of finding the second ion in the state $\ket{0}$ may deviate from the predictions of Eqs. \ref{res_Copenhagen} and \ref{MWI}. To isolate and correct for such gate-induced errors, one can remove the unconscious observer from the setup and directly measure this probability. If all gates were ideal, Eq.~\ref{scalarproduct} would yield $s = \frac{1}{2}$, and the probability of detecting the second ion in the state $\ket{0}$ would be exactly 1. That is, we would expect to always measure the second ion in $\ket{0}$.

However, this probability will fall below 1 if the gates are imperfect. By quantifying the deviation, we can either calibrate and improve the gate fidelities or apply a correction to Eqs.~\ref{res_Copenhagen} and~\ref{MWI} to account for the systematic error. Likewise, statistical uncertainties in the case without the unconscious observer can be evaluated and used to refine the analysis.

An imperfect unconscious observer may occasionally make errors, misidentifying the state of the first qubit. Specifically, it might register $\ket{0}$ when the actual state is $\ket{1}$, and vice versa. In the concrete implementation of the unconscious observer discussed in Section~\ref{sec:example}, this corresponds to sporadic flips of the first ion. Let us denote the probability of such a misidentification—either $\ket{0} \to \ket{1}$ or $\ket{1} \to \ket{0}$—as $p_{\text{err}}$, and assume that $p_{\text{err}} \ll p_1(0)$.

Here are the pre-rotation states for an imperfect unconscious observer and their probabilities with Copenhagen interpretation:
\begin{enumerate}
\item{$\ket{00}$ with probability $p_1(0) - p_{\text{err}}$}
\item{$\ket{01}$ with probability $p_{\text{err}}$}
\item{$\ket{11}$ with probability $p_1(1) - p_{\text{err}}$}
\item{$\ket{10}$ with probability $p_{\text{err}}$}
\end{enumerate}

To get the probability of finding the second qubit in the state $\ket{0}$ post-rotation, we need to replace Eq.~\ref{Copenhagen} with:
\begin{equation}
\begin{split}
p_2(0)_{\text{rot}}^{Copenhagen} = 
(p_1(0)  - p_{err})\times \left| \frac{1}{\sqrt{2}} \hat{P}_2^0 ( \ket{00} + \ket{11})  \right|^2 + \\
p_{err}\times \left| \hat{P}_2^0 ( \ket{01})  \right|^2 + \\
(p_1(1)  - p_{err})\times \left| \frac{1}{\sqrt{2}} \hat{P}_2^0 ( \ket{00} - \ket{11})  \right|^2 + \\
p_{err}\times \left| \hat{P}_2^0 ( \ket{10})  \right|^2 = \\
\frac{1}{2}((p_1(0)-p_{err}) + \frac{1}{2} (p_1(1)  - p_{err})) + p_{err} = \\
\frac{1}{2} (p_1(0) + p_2(0)) = \frac{1}{2},
\end{split}
\end{equation}

which matches Eq. \ref{res_Copenhagen}. 

In the unitary evolution interpretations, the pre-rotation state of the system is:
\begin{equation}
    \begin{split}
        \ket{\psi'} = 
        \sqrt{p_1(0)  - p_{err}}\ket{00}\ket{obs'_{00}} + \\
        \sqrt{p_{err}}\ket{01}\ket{obs'_{01}} + \\
        \sqrt{p_1(0)  - p_{err}}\ket{11}\ket{obs'_{11}} + \\
        \sqrt{p_{err}}\ket{10}\ket{obs'_{10}},
    \end{split}
\end{equation}
where all phases are absorbed into the observers states: $\ket{obs'_{00}}$, $\ket{obs'_{01}}$, $\ket{obs'_{10}}$, $\ket{obs'_{11}}$.
After rotation:
\begin{equation}
    \begin{split}
        \ket{\psi'_{rot}} = 
        \sqrt{p_1(0)  - p_{err}}\frac{1}{\sqrt{2}}(\ket{00} + \ket{11})\ket{obs'_{00}} + \\
        \sqrt{p_{err}}\ket{01}\ket{obs'_{01}} + \\
        \sqrt{p_1(0)  - p_{err}}\frac{1}{\sqrt{2}}(\ket{00} - \ket{11})\ket{obs'_{11}} + \\
        \sqrt{p_{err}}\ket{10}\ket{obs'_{10}},
    \end{split}
\end{equation}

The probability to find the second qunit in state $\ket{0}$ then is:
\begin{equation}
    \begin{split}
        p_2(0)^U_{rot} = \braket{\hat{P}_2^0  \psi'_{rot}}{\hat{P}_2^0  \psi'_{rot}} = \\
        \frac{1}{2}(p_1(0)-p_{err})
        (\bra{obs'_{00}} + \bra{obs'_{11}})(\ket{obs'_{00}} + \ket{obs'_{11}}) + p_{err} = \\
        (p_1(0)-p_{err})(1 +Re(\braket{obs'_{00}}{obs'_{10}}) + p_{err} \sim \\
        \frac{1}{2}(1 + Re(\braket{obs'_{00}}{obs'_{11}})),
    \end{split}
\end{equation}
which matches Eq. \ref{MWI}. This result confirms that, even accounting for small observer errors, the predicted measurement probability remains consistent with the Copenhagen interpretation.

If the errors are asymmetric, we can adjust Eq. \ref{rotation} and modify the experiment to mitigate the errors. If the probability for $\ket{0}$ to be misidentified as $\ket{1}$ is $p_{\text{err}}^{\uparrow}$, and the probability for $\ket{1}$ to be misidentified as $\ket{0}$ is $p_{\text{err}}^{\downarrow}$, the rotational transformation takes the following form:
\begin{equation} \label{complex_rotation}
\begin{split}
\ket{00} \to (\alpha_{\text{rot}}\ket{00}+\beta_{\text{rot}}\ket{11}) \\
\ket{11} \to (\beta_{\text{rot}}\ket{00}-\alpha_{\text{rot}}\ket{11}),
\end{split}
\end{equation} where
\begin{equation}
\left|\alpha_{\text{rot}}\right|^2 = \frac{1}{2}\frac{1-2p_{\text{err}}^{\downarrow}}{1-p_{\text{err}}^{\uparrow}-p_{\text{err}}^{\downarrow}},
\end{equation}
and
\begin{equation}
\left|\alpha_{\text{rot}}\right|^2 + \left|\beta_{\text{rot}}\right|^2 = 1.
\end{equation}

Consequently, asymmetric errors also do not alter our conclusion: experimenters can still differentiate between the interpretations.
 
\section*{Appendix C. Shift in the location of QD}
Let's evaluate how much will the QD move after the interaction with the first ion. We will represent the QD-ion interaction as an interaction with a potential well. Due to the conservation of momentum, $x$ component of the first ion's momentum changes after the measurement.

Let's define $ f $ as the ratio of the average passed ion energy $ <E_{\text{passed}}^{\text{ion}}> $ to the initial electron energy $ <E_{\text{initial}}^{\text{ion}}> $:
\begin{equation}
f = \frac{<E_{\text{passed}}^{\text{ion}}>} {<E_{\text{initial}}^{\text{ion}}>}
\end{equation}

This implies that the QD will recoil such that $<{p'_x}^{QD}> = <{p_x}^{\text{ion}}> - <{p_x}^{\text{passed}}>$, where $ <{p'_x}^{QD}> $, 
$ <p_x^{\text{ion}}> $, and $ <p_x^{\text{passed}}> $ are the average momenta of the QD, the ion before interaction, and the traversed ion, respectively.

Thus, the average momentum of the QD after the measurement is given by 
\begin{equation}
    <p'_{x^{QD}}> = <p_x^{\text{ion}}> - <p_x^{\text{passed}}> =\sqrt{2 m^{\text{ion}} <E_{\text{initial}}^{\text{ion}}> ( 1 - f )}, 
\end{equation} where $ m^{\text{ion}} $ is the mass of the first ion.

Consequently, if an ion passes through the upper slit, the average momentum of the QD will be
\begin{equation}
<{p'_x}^{QD}> = \sqrt{2 m^{\text{ion}} <E_{\text{initial}}^{\text{ion}}> ( 1 - f )},
\end{equation}
and zero otherwise. 

For $ f = 1\% $, 1000 atoms in a QD, and $ E^{\text{ion}}_{\text{initial}} = 10^{-4} \, \text{eV} $, the velocity of the QD after the interaction is approximately $ 0.1\text{mm/s} $.

\end{document}